\documentclass{PoS}
\usepackage{cite,amsmath}
\pdfoutput=1

\title{Constraints on New Physics from $B$ mesons}

\ShortTitle{Constraints on New Physics from $B$ mesons}

\author{\speaker{Monika Blanke}\\
Institute for Nuclear Physics (IKP) and Institute for Theoretical Particle Physics (TTP), Karlsruhe Institute of Technology (KIT), D-76128 Karlsruhe, Germany\\
        E-mail: \email{monika.blanke@kit.edu}}

\abstract{These proceedings review the status of New Physics contributions to flavour violating $B$ decays. The anomalies in charged and neutral current $B$ decays related to lepton flavour universality violation have received a substantial amount of attention over the past years, and we discuss the current status in light of the new data presented earlier this year. We also recall a tension in the neutral $B$ meson mixing observables $\Delta M_d$ and $\Delta M_s$ and in particular their ratio, when compared with their SM predictions obtained using tree-level determinations of the CKM matrix and the recent lattice QCD results for the relevant hadronic matrix elements.
Last but not least, we advocate  kaon physics as a unique probe of very high energy scales and briefly discuss the current status of $\varepsilon'/\varepsilon$ and $K\to\pi\nu\bar\nu$.\\

Preprint numbers: TTP19-027, P3H-19-029, INT-PUB-19-038}

\FullConference{XXIX International Symposium on Lepton Photon Interactions at High Energies - LeptonPhoton2019\\
		August 5-10, 2019\\
		Toronto, Canada}

\begin{document}


\section{Introduction}

In spite of the convincing arguments that led us to expect the presence of New Physics (NP) at the TeV scale, still no new particles beyond the Standard Model (SM) have been discovered at the LHC and the data on electroweak precision constraints and Higgs physics are in impressive agreement with the SM. It thus appears that the new particles are either too heavy or too weakly coupled to leave a visible imprint on these observables. Barring the construction of a future ultra high energy collider, the search for NP thus requires indirect methods that are even more sensitive to small NP contributions than the ones listed above.

A prime opportunity in this respect is provided by flavour physics, more specifically by flavour changing neutral current (FCNC) processes. Due to their loop-, CKM- and GIM-suppression, they are genuinely small in the SM and thus offer an excellent test of the presence of NP, probing scales far beyond the TeV regime.

In order to exploit the full capacity of FCNCs to  explore physics beyond the SM, high precision is needed both in experimental measurements of flavour violating decays and in theoretical predictions of the SM contributions. For the latter, perturbative and non-perturbative contributions and the precise knowledge of the relevant input parameters, such as the CKM elements, are crucial.


\section{CKM determinations and New Physics in neutral $B$ meson mixing}

To obtain accurate determinations of the elements of the CKM matrix, it is desirable to measure flavour changing charged currents decays, as those are mediated by tree-level exchanges of the $W$ boson in the SM and therefore insensitive to NP contributions. Having in this way fully determined the CKM matrix by the measurement of four independent parameters, one is prepared to make predictions for the SM contributions to FCNC processes that are potentially affected by NP.

At present, unfortunately, a precise determination of the CKM matrix solely through tree-level decays is not possible. While the element $|V_{us}|$ is known to very good accuracy, the tensions between inclusive and exclusive determinations of $|V_{ub}|$ \cite{Bouchard:2019all} and to a lesser extent also $|V_{cb}|$ \cite{Gambino:2019sif}
persist and lead to a significant uncertainty in the determination of the length of the side $R_b$ of the unitarity triangle. Therefore, in order to obtain a precise result for the full Unitarity Triangle (UT), one still needs to rely on the measurement of $\sin2\beta$ from the time-dependent CP asymmetry in $B \to J/\psi K_S$. The latter is loop-induced in the SM and therefore potentially affected by NP. 

\begin{figure}
\includegraphics[width=.61\textwidth]{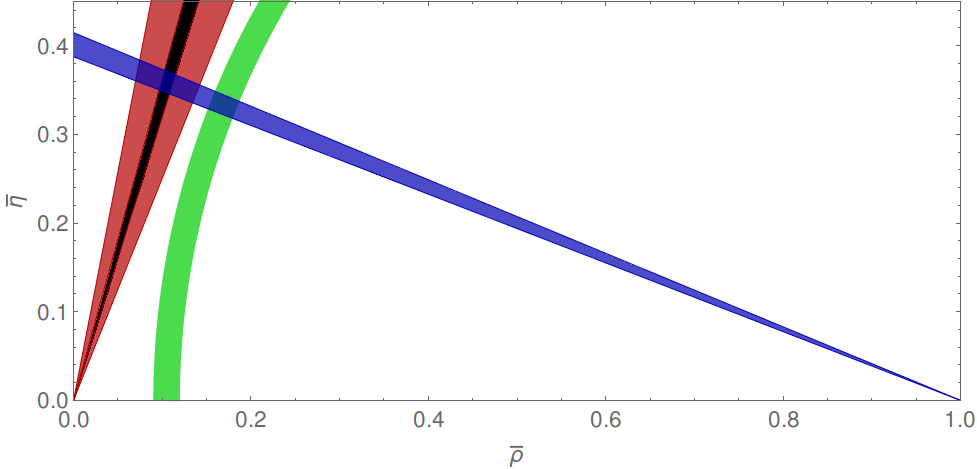}\hfill
\includegraphics[width=.36\textwidth]{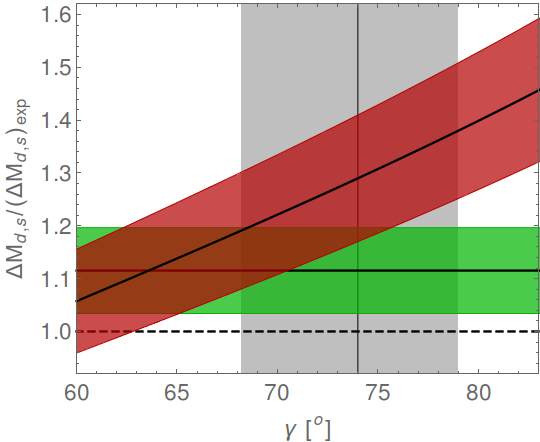}
\caption{\label{fig1}Left: Constraints on the Unitarity Triangle from the measurement of $\sin2\beta$ (blue), the ratio $\Delta M_d/\Delta M_s$ (green), and the tree-level determination of the angle $\gamma$ (red). The future expected $1^\circ$ sensitivity for $\gamma$ by LHCb and Belle II is shown in black. 
Right: SM predictions for $\Delta M_d$ (red) and $\Delta M_s$ (green), normalised to their experimental values, as a function of $\gamma$. The LHCb measurement of the latter angle is displayed by the grey band.
Figures taken from \cite{Blanke:2018cya}.
}
\end{figure}

The determination of the angle $\gamma$ in $B\to DK$ decays, on the other hand, has substantially improved by the LHCb collaboration, providing $\gamma = (74.0^{+5.0}_{-5.8})^\circ$ \cite{Kenzie:2319289}. In addition, the prospects for further improvements are excellent: the expected sensitivity at both LHCb and Belle~II  is at the $1^\circ$ level \cite{Cerri:2018ypt,Kou:2018nap}, with negligible theoretical uncertainties \cite{Brod:2013sga}. In the left panel of Figure \ref{fig1} \cite{Blanke:2018cya} the measured value of $\gamma$ is shown in red, while the $\sin2\beta$ constraint \cite{Amhis:2016xyh} is given in blue. Already now, the resulting determination of the apex of the UT is quite precise, and it will improve considerably with future measurements of $\gamma$, indicated in black.

Comparing this result to the determination of the side $R_t$ through the ratio of mass differences $\Delta M_d/\Delta M_s$ in the $B_{d,s}-\bar B_{d,s}$ meson systems, shown in green, we observe a tension between the direct measurement of $\gamma$ and its indirect determination at the $2\sigma$ level. Using the Fermilab/MILC result \cite{Bazavov:2016nty} for the ratio $\xi$ of hadronic matrix elements entering $B_{d,s}$ mixing, the latter yields $\gamma = (63.0\pm 2.1)^\circ$ \cite{Blanke:2016bhf}, lower than the direct measurement by almost $2\sigma$ \cite{Blanke:2018cya}.  Using instead the values of $\xi$ found by RBC/UKQCD \cite{Boyle:2018knm}, HPQCD \cite{Dowdall:2019bea}, or QCD sum rules \cite{Grozin:2016uqy,Grozin:2017uto,Kirk:2017juj,Grozin:2018wtg,King:2019lal} yields similar results. This tension, if confirmed by future more accurate determinations of $\gamma$, would unambiguously imply the presence of NP in $\Delta M_d$ and/or $\Delta M_s$. 

To get a better picture of the NP underlying this tension, the right panel of Figure \ref{fig1} shows the SM predictions of $\Delta M_d$ and $\Delta M_s$ as functions of the angle $\gamma$ \cite{Blanke:2018cya}. In making these predictions, again the hadronic matrix elements provided by Fermilab/MILC  \cite{Bazavov:2016nty} were used, which dominate the current FLAG averages \cite{Aoki:2019cca}. We find that the SM prediction for $\Delta M_d$ exceeds its experimental value by about 30\%, so that a significant negative NP contribution is required. Also $(\Delta M_s)_\text{SM}$ appears to be somewhat above the data, but the tension is less severe. 

This pattern of deviations, if eventually confirmed, would imply the presence of flavour non-universal NP contributions to $\Delta F=2$ transitions, with the effects being larger in $b\to d$ than in $b\to s$ transitions. This pattern can neither be accommodated within Constrained Minimal Flavour Violation \cite{Buras:2000dm,Buras:2003jf,Blanke:2006ig} nor in  models with a minimally broken $U(2)^3$ flavour symmetry \cite{Kagan:2009bn,Barbieri:2011ci,Barbieri:2012uh,Buras:2012sd}, but calls for the presence of new sources of flavour violation, unless new operators beyond the SM $(V-A)\otimes(V-A)$ one are present. 
The required destructive interference between the SM and NP contributions is particularly interesting, as it can most easily be generated with the help of a large CP-violating phase $\sim\pi/2$ governing the $b\to d$ transition, resulting in a CP-phase $\sim\pi$ in the corresponding $\Delta F =2$ mode \cite{Blanke:2009pq}. The hint for an anomaly in $\Delta M_d$ therefore calls for non-standard effects in rare and CP-violating decays governed by the $b\to d$ current  \cite{Blanke:2018cya}.

The presence of an anomaly in $\Delta M_d$ is however not unambiguous. The very recent HPQCD results \cite{Dowdall:2019bea} for the hadronic matrix elements governing $\Delta B =2$ transitions, based on 2+1+1 dynamical quark flavours and using a different method for extracting the continuum limit than done by Fermilab/MILC \cite{Bazavov:2016nty}, 
show no deviation from the data in either $\Delta M_d$ or $\Delta M_s$, so that the origin of the tension in $\gamma$ remains unresolved. Also QCD sum rules do not come at a rescue here, since their recent results \cite{King:2019lal} are compatible with both Fermilab/MILC and HPQCD determinations. It remains to be seen what lattice QCD will eventually tell us about the size of the relevant hadronic matrix elements.

Finally note that lowering $|V_{cb}|$ from the inclusive value used here softens the individual tensions in $\Delta M_d$ and $\Delta M_s$, but has no impact on the discrepancy between the tree-level value of $\gamma$ and $\Delta M_d/\Delta M_s$, and in addition introduces a tension in the parameter $\epsilon_K$ describing CP violation in neutral kaon mixing \cite{Blanke:2018cya,Blanke:2016bhf,Bailey:2018feb}.


\section{The $R(D^{(*)})$ anomaly}

Over the past years, several deviations from the SM in lepton flavour universality violating $B$ decays emerged and have attracted a lot of attention in the theory community. Several years ago, the $B$-factories BaBar \cite{Lees:2012xj,Lees:2013uzd} and Belle \cite{Huschle:2015rga,Hirose:2016wfn,Hirose:2017dxl} found the ratios
\begin{equation}
R(D^{(*)})=\frac{\text{BR}(B\to D^{(*)} \tau
  \nu)}{\text{BR}(B\to D^{(*)} \ell \nu)} \qquad (\ell=e,\mu)
\end{equation}
significantly above their SM predictions. More recently, LHCb contributed by providing measurements of $R(D^*)$ \cite{Aaij:2015yra,Aaij:2017uff,Aaij:2017deq} and $R(J/\psi)$ \cite{Aaij:2017tyk}, also hinting for the presence of an anomaly. With the most recent analysis of $R(D)$ and $R(D^*)$ by Belle \cite{Abdesselam:2019dgh}, the tension with the SM was somewhat reduced, but is still found at the $3.1\sigma$ level \cite{Amhis:2016xyh}.

Interestingly, a model-independent sum rule \cite{Blanke:2018yud} relates $R(D)$ and $R(D^*)$ to the baryonic ratio
\begin{equation}
R(\Lambda_c)=\frac{\text{BR}(\Lambda_b \to \Lambda_c \tau
  \nu)}{\text{BR}(\Lambda_b \to \Lambda_c \ell \nu)} \qquad (\ell=e,\mu)
\end{equation}
that can be measured by LHCb and will provide an experimental consistency check of the anomaly. The current prediction reads \cite{Blanke:2019qrx}
\begin{equation}
R(\Lambda_c) = R_\text{SM}(\Lambda_c)(1.15\pm 0.04)
=0.38\pm0.01\pm0.01\,.
\end{equation}

From the theoretical point of view, the underlying $b\to c\tau\nu$ transition is conveniently described by the effective Hamiltonian
\begin{equation}
 {\cal H}_{\rm eff}=  2\sqrt{2} G_{F} V^{}_{cb} \big[(1+C_{V}^{L}) O_{V}^L +   C_{S}^{R} O_{S}^{R} 
   +C_{S}^{L} O_{S}^L+   C_{T} O_{T}\big] 
\label{Heff}
\end{equation}
with the Wilson coefficients $C_i$ describing the NP contribution.\footnote{For notations and conventions, and phenomenological expressions of the relevant observables in terms of Wilson coefficients, see \cite{Blanke:2018yud}.}

Several simplified NP models, in which the $b\to c\tau\nu$ transition is induced by the tree-level exchange of a single new heavy mediator, have been discussed in the literature. Possible contributions from a heavy charged $W'$ gauge boson, leading to a change $C_V^L\ne 0$ of the SM $(V-A)\otimes (V-A)$ current, have been put forward in \cite{He:2012zp,Greljo:2015mma}, but they are challenged by high-$p_T$ di-$\tau$ data at the LHC \cite{Faroughy:2016osc}. Other possibilities are the exchange of a charged Higgs boson \cite{Kalinowski:1990ba,Hou:1992sy,Crivellin:2012ye} or of a leptoquark, with various spin and coupling structures possible, see e.\,g.\ \cite{Freytsis:2015qca}.

\begin{figure}
\centering{\includegraphics[width=.8\textwidth]{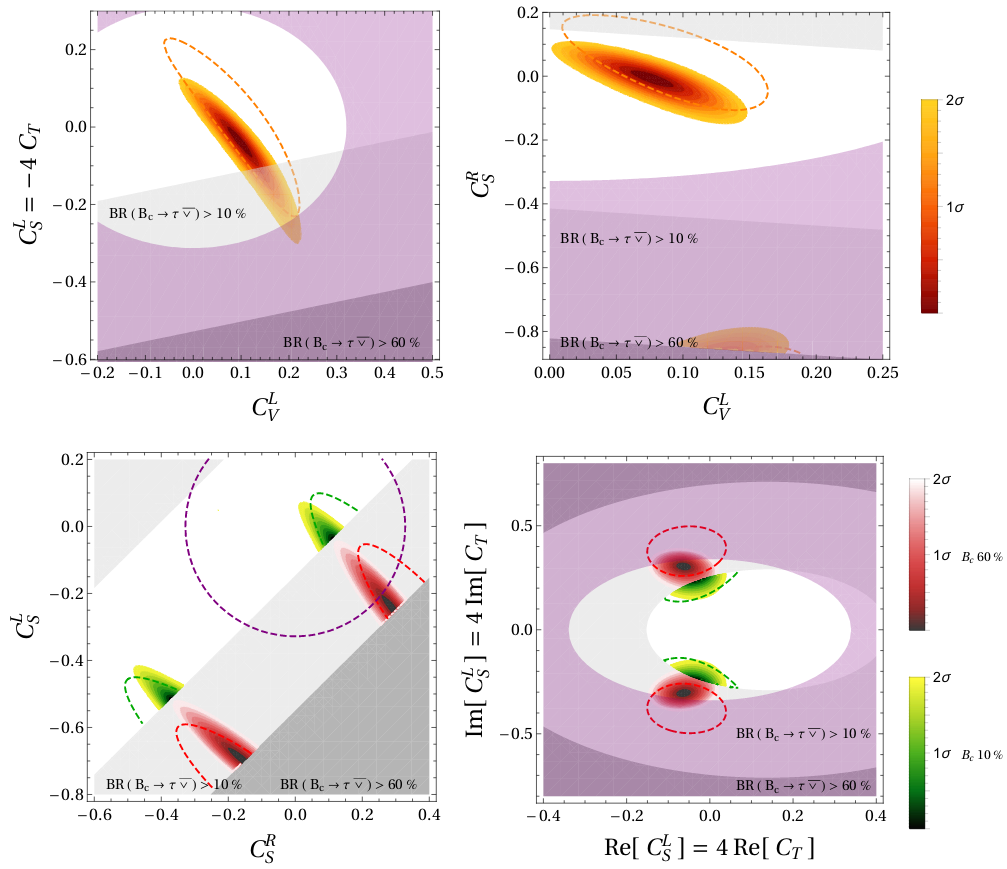}}
\caption{Fit results for various two-dimensional scenarios of NP in $b\to c\tau\nu$. The grey-shaded areas display the contribution to the branching ratio $\text{BR}(B_c\to\tau\nu)$, while the constraints from LHC mono-$\tau$ searches are shown in purple.
Figures taken from \cite{Blanke:2019qrx}.
\label{fig2}}
\end{figure}

The current situation of charged Higgs and leptoquark scenarios is shown in Figure \ref{fig2} \cite{Blanke:2019qrx}.\footnote{For recent global fit results, see also \cite{Murgui:2019czp,Shi:2019gxi,Aebischer:2019mlg}.} Both the $SU(2)_L$-singlet scalar leptoquark \cite{Deshpande:2012rr,Tanaka:2012nw,Sakaki:2013bfa}, shown in the upper left plot, and the $SU(2)_L$-singlet vector leptoquark \cite{Alonso:2015sja,Calibbi:2015kma,Fajfer:2015ycq,Bordone:2017bld}, see upper right plot, provide a good fit to the available $b\to c\tau\nu$ data, with modest contributions to the $B_c\to\tau\nu$ branching ratio (displayed in grey) and in agreement with the high-$p_T$ constraints from mono-$\tau$ searches (purple region excluded) \cite{Greljo:2018tzh}. The scalar $SU(2)_L$-doublet leptoquark (lower right plot) yields a good fit to the data only if its couplings are allowed to be complex, i.\,e.\ CP-violating \cite{Becirevic:2018afm}. In this scenario a significant contribution to $\text{BR}(B_c\to\tau\nu)\sim 20\%$ is predicted, and the best-fit point is on the verge of being tested by the mono-$\tau$ searches. 
The best fit to the low-energy $b\to c\tau\nu$ data is currently provided by the charged Higgs scenario shown in the lower left figure, as only in this case, the measurement of the $D^*$ polarisation $F_L(D^*)$ \cite{Abdesselam:2019wbt} can be accommodated within $1\sigma$. Note however that the best-fit point in this case is in tension with the mono-$\tau$ data, and a large branching ratio $\text{BR}(B_c\to\tau\nu)>50\%$ is predicted. While the latter has not been measured directly, upper bounds of 30\% \cite{Alonso:2016oyd} and even 10\% \cite{Akeroyd:2017mhr} have been put forward in the literature. A recent critical reassessment however showed that even values as large as 60\% cannot be excluded at present \cite{Blanke:2018yud,Blanke:2019qrx}.

Complementary information on the NP model at work can be obtained from the measurement of differential and angular observables \cite{Nierste:2008qe,Becirevic:2016hea,Celis:2016azn,Iguro:2018vqb,Blanke:2018yud,Becirevic:2019tpx}, such as the $D^*$ and $\tau$ polarisations $F_L(D^*)$ and $P_\tau (D^{(*)})$, whose correlations turn out to discriminate well between the different scenarios. In addition to precise measurements, which appear to be rather challenging, also a better theoretical understanding of the underlying form factors is needed to fully exploit their potential.

Finally let us stress that already now stringent constraints on the various NP scenarios arise when taking into account relations implied by the electroweak $SU(2)_L$ symmetry. The latter implies potentially large contributions to decays like $B\to K^{(*)}\nu\bar\nu$, $B_s\to\tau^+\tau^-$ and $B\to K^{(*)}\tau^+\tau^-$ \cite{Calibbi:2015kma,Crivellin:2017zlb}. Similarly significant rates for 
$\Upsilon\to\tau^+\tau^-$ or $\psi\to\tau^+\tau^-$ are expected \cite{Aloni:2017eny}. All in all, while not excluded at present, a full resolution of the $R(D^{(*)})$ anomaly in terms of a UV-complete model appears to be challenging.


\section{Anomalies in $b\to s\ell^+\ell^-$ transitions}

The other set of $B$ decay anomalies that received significant attention in the theory community is related to the semileptonic $b\to s\ell^+\ell^-$ transitions. Most noticeable in this respect are a $3.4\sigma$ deviation in the angular analysis of the $B\to K^*\mu^+\mu^-$ decay \cite{Aaij:2015oid}, as well as the suppression of the lepton flavour universality ratios
\begin{equation}
R_{K^{(*)}} = \frac{\text{BR}(B\to K^{(*)}\mu^+\mu^-)}{\text{BR}(B\to K^{(*)}e^+e^-)}
\end{equation}
below unity with more than $2\sigma$ significance in various  $q^2$ bins \cite{Aaij:2017vbb,Aaij:2019wad}, all found by the LHCb Collaboration. Measurements of some of these quantities by Belle, ATLAS and CMS exist, however their uncertainties are currently too large to be conclusive. For a complete review of the current experimental situation, we refer the reader to \cite{bsll-status}.

For a model-independent description of non-SM effects, again the effective theory description is most suitable. The effective Hamiltonian for $b\to s\ell^+\ell^-$ transition reads
\begin{equation}
\mathcal{H}_\text{eff}= -\frac{4 G_F}{\sqrt{2}} V_{tb}^* V_{ts} \frac{e^2}{16\pi^2}\sum_i(C_i {\cal O}_i +C'_i {\cal O}'_i)+h.c.\,,
\end{equation}
with the electromagnetic dipole operators $\mathcal{O}_7$ and $\mathcal{O}'_7$ and the semileptonic current-current operators $\mathcal{O}_9$, $\mathcal{O}'_9$, $\mathcal{O}_{10}$ and $\mathcal{O}'_{10}$ being most sensitive to NP.\footnote{The (pseudo)scalar operators are strongly constrained by $\text{BR}(B_s\to\mu^+\mu^-)$ and therefore do not yield a relevant contribution to the semileptonic decays.} Note that the semileptonic operators are loop-suppressed in the SM, but can be mediated by tree-level transitions in the presence of NP.

\begin{figure}
\centering{\includegraphics[width=.55\textwidth]{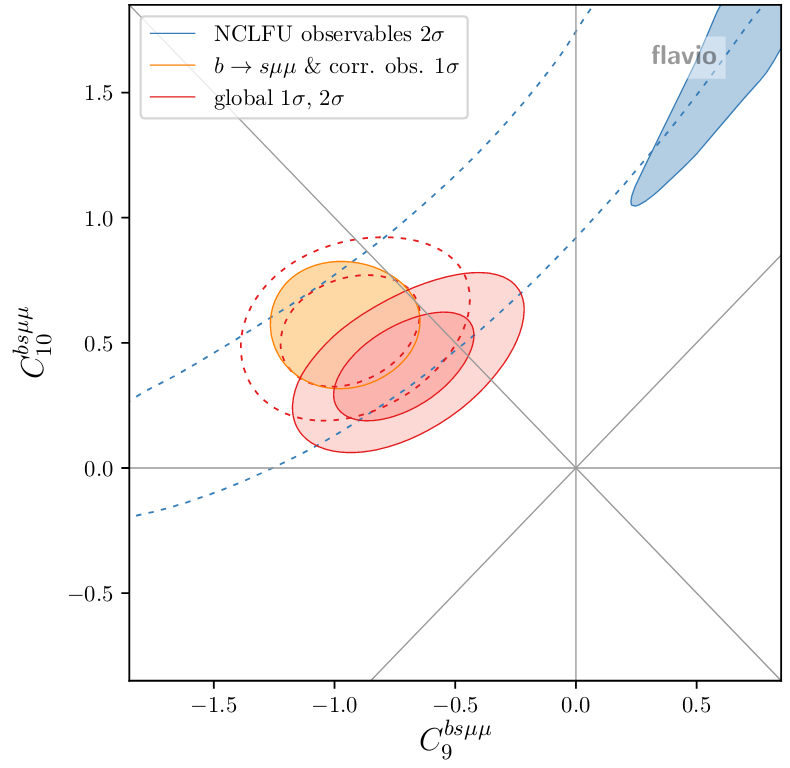}}
\caption{
Result of a global fit (red) to $b\to s\ell^+\ell^-$ observables in the $\left(C_9^{bs\mu\mu},C_{10}^{bs\mu\mu}\right)$ plane, assuming all other Wilson coefficients to be SM-like. The constraint from the LFU ratios is shown in blue, while the fit to $b\to s\mu^+\mu^-$ data is displayed in orange.
Figure taken from \cite{Aebischer:2019mlg}.
\label{fig3}
}
\end{figure}

Several groups have performed global fits including the data presented last spring \cite{Aebischer:2019mlg,Alguero:2019ptt,Arbey:2019duh}. The differences in the results between the various groups are marginal, and we follow here the presentation in \cite{Aebischer:2019mlg} which found the main results:
\begin{itemize}
\item
The 1D scenarios providing the best overall agreement with the data are $C_9^{bs\mu\mu} \simeq -0.97$ and  $C_{9}^{bs\mu\mu}=-C_{10}^{bs\mu\mu} \simeq -0.53$, with pulls of about $6\sigma$ relative to the SM.
\item
A non-zero $C_{10}^{bs\mu\mu}$ is required to accommodate the suppression of $\text{BR}(B_s\to \mu^+\mu^-)$ with respect to its SM value, as indicated by the ATLAS \cite{Aaboud:2018mst}, CMS \cite{CMS:2019qnb} and LHCb \cite{Aaij:2017vad} data.
\item
As can be seen in Figure \ref{fig3}, the fits to either the $b\to s\mu^+\mu^-$ data or to the LFU ratios are not fully consistent if only NP in the muon channel is assumed (see also \cite{Alguero:2018nvb}). Note however that this does not necessarily require NP to couple directly to electrons -- a non-zero $C_9^{bsee}$ can also be generated radiatively from NP in the $b\to s\tau^+\tau^-$ channel \cite{Crivellin:2018yvo}.
\end{itemize}

Turning our attention from the EFT picture to concrete NP models, the most popular explanations include the tree-level exchange of a heavy $Z'$ gauge boson \cite{Altmannshofer:2013foa,Gauld:2013qja,Altmannshofer:2014cfa,Crivellin:2015lwa,Descotes-Genon:2017ptp}, loop-induced box contributions \cite{Gripaios:2015gra,Arnan:2016cpy,Arnan:2019uhr} or $Z'$ penguins \cite{Belanger:2015nma,Kamenik:2017tnu}, or a tree-level leptoquark contribution \cite{Hiller:2014ula,Alonso:2015sja,Fajfer:2015ycq,Calibbi:2015kma,Becirevic:2016yqi}.

For the sake of brevity, we do not attempt to provide a comprehensive review of the NP model landscape addressing the $b\to s\ell^+\ell^-$ anomalies. Instead we content ourselves with a few remarks on what currently appears to be the most popular NP explanation: the $SU(2)_L$-singlet vector leptoquark. From the experimental side, this simplified model is  least constrained by complementary observables such as $B_s- \bar B_s$ mixing or $B\to K^{(*)}\nu\bar\nu$. From the phenomenological point of view, this leptoquark can offer a combined solution to both the $b\to s\ell^+\ell^-$ and the $b\to c\tau\nu$ anomalies. And last but not least, the $SU(2)_L$-singlet vector leptoquark is also theoretically appealing since it naturally arises as a heavy degree of freedom from the spontaneously broken Pati-Salam gauge symmetry
\begin{equation}
SU(4)_c \times SU(2)_L \times SU(2)_R
\end{equation}
unifying quarks and leptons \cite{Pati:1974yy}. 

This observation offers an interesting perspective on building a UV-completion of the simplified model extending the SM by just one leptoquark state with the required coupling structures introduced by hand. The main challenge for model-builders is then to generate flavour non-universal couplings of the leptoquark that are necessary to allow for its existence at the TeV scale despite the stringent constraints from $K_L\to\mu e$ and $K\to\pi\mu e$ \cite{Hung:1981pd,Valencia:1994cj}, and to accommodate the observed pattern of effects in $b\to c$ and $b\to s$ transitions including lepton flavour universality violation.

The required flavour pattern can be obtained from the mixing of the SM fermions with heavy vectorlike fermions, either introduced explicitly as new heavy degrees of freedom \cite{DiLuzio:2017vat,Calibbi:2017qbu,Bordone:2017bld,Greljo:2018tuh,Balaji:2018zna}, or resulting as the massive Kaluza-Klein states of a compactified extra dimension \cite{Barbieri:2016las,Blanke:2018sro}.\footnote{The model introduced in \cite{Barbieri:2016las} can be considered as the 4D strongly coupled dual of a 5D warped model.} Interestingly, by symmetry principles, all of these models introduce a heavy spin-1 colour octet in the same mass range as the leptoquark that is stringently constrained by direct searches at the LHC \cite{Cornella:2019hct}.


\section{The discovery potential of kaon decays}

Last but not least, let us address the question in which flavour-violating observables large NP effects are generally expected. As explained earlier in these proceedings, significant NP contributions to flavour-changing charged current decays, like $b\to c\tau\nu$, are relatively hard to accommodate due to the significant SM contributions. Sizeable NP effects in FCNC observables like $B_{d,s}-\bar B_{d,s}$ mixing or the semileptonic $b\to s$ transitions, are more natural in the sense that they have to compete only with a highly suppressed loop-induced SM effect. 

In order to understand in which meson system NP effects can be most pronounced, it is useful to have a look at the hierarchical structure of the CKM matrix, as the latter governs the relative size of SM contributions. While FCNC observables related to $b\to d$ and $b\to s$ transitions are governed by $V_{tb}^* V_{td} \sim  10^{-2}$ and $V_{tb}^* V_{ts} \sim  4\cdot 10^{-2}$, respectively, $s\to d$ transitions receive a much stronger suppression\footnote{Effects from virtual charm quarks, instead, are GIM-suppressed by the smallness of the charm quark mass.}  by $V_{ts}^* V_{td} \sim 5\cdot 10^{-4}$. FCNC transitions are therefore significantly smaller in the kaon sector and hence intrinsically more sensitive even to small NP effects than the corresponding $B$ decay observables.

\begin{figure}
\centering{\includegraphics[width=.4\textwidth]{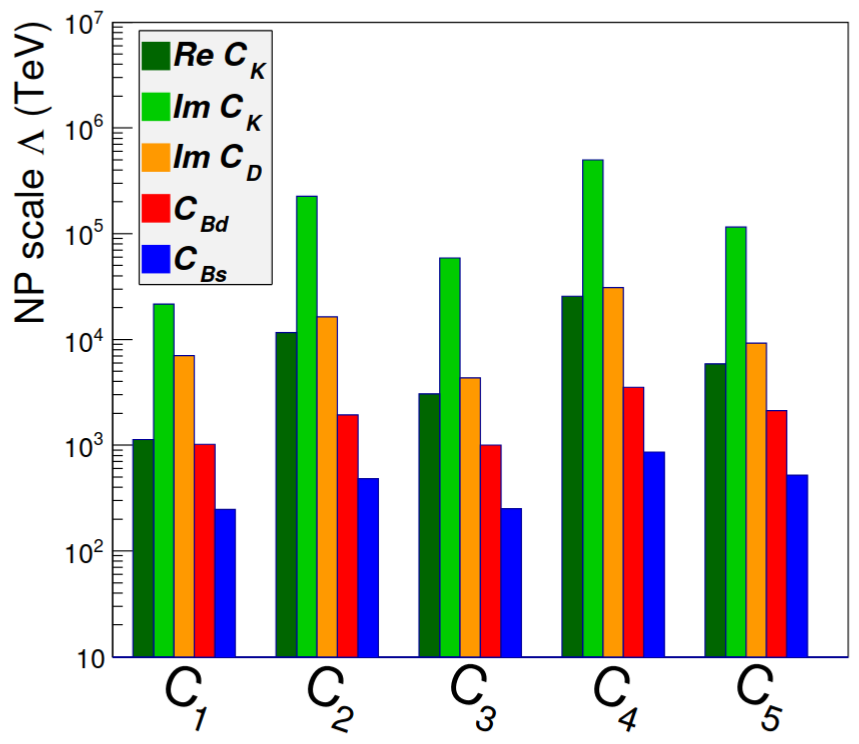}\qquad
\includegraphics[width=.48\textwidth]{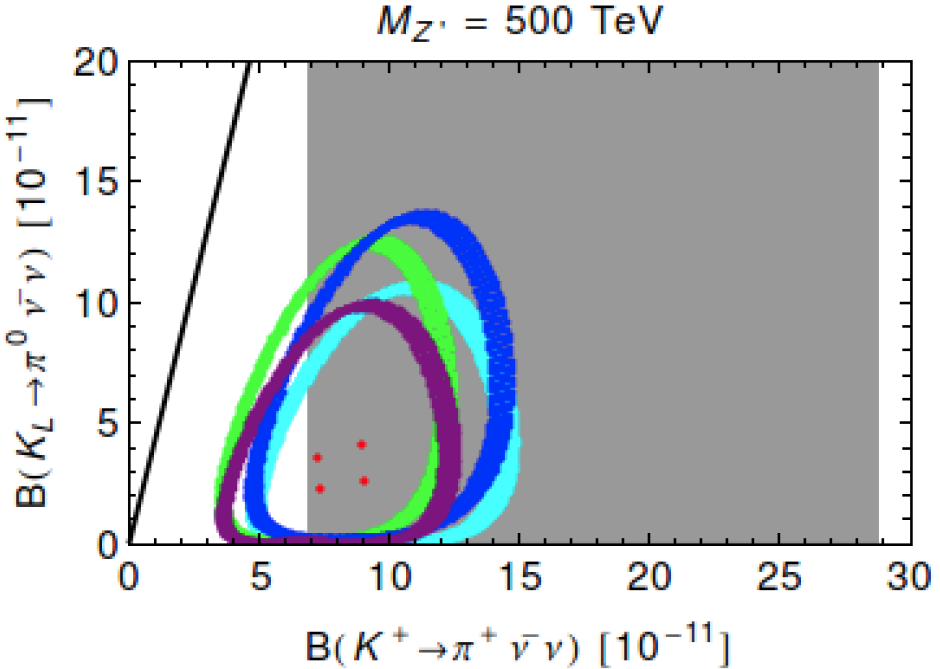}}
\caption{Left: Constraints on the scale of NP of the various operators contributing to neutral meson mixing. Figure taken from \cite{Bona:2017cxr}. Right: The rare decay branching ratios $\text{BR}(K^+\to\pi^+\nu\bar\nu)$ and $\text{BR}(K_L\to\pi^0\nu\bar\nu)$ in the presence of a heavy flavour-violating $Z'$ gauge boson with mass $M_{Z'}=500\,\text{TeV}$. Figure taken from \cite{Buras:2014zga}.
\label{fig4}}
\end{figure}

This expectation is confirmed by the constraints on the scale of NP deduced from the experimental data on neutral meson mixing observables, as shown in the left panel of Figure \ref{fig4} \cite{Bona:2017cxr}. The constraints assume the absence of any suppression mechanism, like loop factors or flavour hierarchies, in the NP sector other than the high scale $\Lambda$. The extremely tiny CP-violating effects in $K^0-\bar K^0$ mixing, measured by $\varepsilon_K$, limit the scale of generic NP contributions to be above $10^5\,\text{TeV}$, while the bounds from $B_{d,s}-\bar B_{d,s}$ mixing are weaker by two orders of magnitude. 

With this picture in mind, the recent hints for a potential anomaly in $\varepsilon'/\varepsilon$ \cite{Bai:2015nea,Buras:2015xba,Buras:2015yba,Kitahara:2016nld,Buras:2018ozh} is not surprising. This parameter measures direct CP-violation in $K\to\pi\pi$ decays and is strongly suppressed in the SM not only by the CKM hierarchy, but in addition by an accidental cancellation between contributions from QCD and electroweak penguins that enter with a relative minus sign. While the experimental determination \cite{Batley:2002gn,AlaviHarati:2002ye,Abouzaid:2010ny}
\begin{equation}
\varepsilon'/\varepsilon = (16.6 \pm 2.3) \cdot 10^{-4}
\end{equation}
has been with us for over 15 years, only in the past years lattice QCD calculations of the relevant hadronic matrix elements became possible \cite{Bai:2015nea}, leading to the SM prediction \cite{Buras:2015yba,Kitahara:2016nld}
\begin{equation}
(\varepsilon'/\varepsilon)_\text{SM} = (1.9\pm 4.5)\cdot 10^{-4}\,,
\end{equation}
which is below the data by $2.9\sigma$. Interestingly the presence of an anomaly is supported by dual QCD calculations \cite{Buras:2015xba,Buras:2016fys}; however it is not seen with the use of chiral perturbation theory methods \cite{Gisbert:2017vvj}. The ball is now in the field of the lattice QCD experts to provide improved determinations of the relevant $K\to\pi\pi$ matrix elements in order to be able to draw definite conclusions. 

A plethora of NP models exist that can rather easily accommodate NP contributions to $\varepsilon'/\varepsilon$ at the level of $10^{-3}$ \cite{Blanke:2015wba,Buras:2015yca,Buras:2015kwd,Kitahara:2016otd,Bobeth:2016llm,Endo:2017ums,Haba:2018byj}. Interestingly, quite generally a NP contribution in  $\varepsilon'/\varepsilon$ is correlated with a deviation of the branching ratio $\text{BR}(K_L\to\pi^0\nu\bar\nu)$ from its SM prediction, as also the latter is sensitive to direct CP violation in the kaon system.

Both $K_L\to\pi^0\nu\bar\nu$ and $K^+\to\pi^+\nu\bar\nu$ are indeed excellent probes of physics beyond the SM, as they are extremely rare and theoretically very clean. Due to their tiny SM rates of the order of $10^{-11}$, they can probe NP at scales well beyond 100\,TeV, as has been shown explicitly in the context of a flavour violating $Z'$ in \cite{Buras:2014zga}, see the right panel of Figure \ref{fig4}. In addition the correlation between the two modes provides insight into the NP operator structure at work in neutral kaon mixing \cite{Blanke:2009pq}.

Fortunately, despite the great challenges to measure these extremely rare decays, the experimental future for the $K\to\pi\nu\bar\nu$ decays is bright. Concerning the charged mode, the NA62 experiment at CERN is currently taking data and already reported the observation of one event in the signal region \cite{CortinaGil:2018fkc}, with an update to be presented soon and a long term goal of determining $\text{BR}(K^+\to\pi^+\nu\bar\nu)$ with 10\% accuracy. For the neutral mode, instead, the 
KOTO experiment at the Japanese facility J-PARC has recently improved the upper limit on the branching ratio \cite{Ahn:2018mvc}, with further data taking and analysis ongoing. Lastly, the recently proposed experiment
KLEVER at CERN \cite{Ambrosino:2019qvz} has the aim of measuring $\text{BR}(K_L\to\pi^0\nu\bar\nu)$ at the 20\% level.


\section{Conclusions}

In the absence of a direct NP discovery at the LHC, flavour physics has attracted an increasing amount of attention over the past years. While this interest is currently mainly triggered by the anomalies in charged and neutral current $B$ decays related to a possible violation of lepton flavour universality, the prospects of observing NP effects in flavour violating transitions reach far beyond these observables. Flavour changing neutral current processes, due to their strong suppression within the SM, are generally expected to be very sensitive even to small NP effects. Indeed a tension seems to be emerging also in neutral $B$ meson mixing data, when compared with the SM predictions using CKM elements determined from tree-level decays. Last but not least, maybe even more exciting is the possible discrepancy in $\varepsilon'/\varepsilon$ implied by the first lattice QCD determination, as rare and CP-violating kaon decays are most sensitive to NP contributions even from very high scales beyond 100\,TeV. In this context also the upcoming improved determinations of the $K\to\pi\nu\bar\nu$ decay rates will be very exciting and may bring pleasant surprises.


\paragraph{Acknowledgements}

I am grateful to the organizers of LP2019 for inviting me to present a talk at this inspiring conference, and for the great effort they made to increase equity and diversity and to ensure that everyone felt welcome and taken care of. 
I acknowledge financial support by the
 Deutsche Forschungsgemeinschaft (DFG, German Research Foundation) under grant  396021762 -- TRR 257, and by the program ``Kongressreisen'' of the Deutscher Akademischer Austauschdienst (DAAD, German Academic Exchange Service). 
 I also thank the Institute for Nuclear Theory at the University of Washington for its hospitality and the Department of Energy for partial support of my participation at the program ``Heavy-Quark Physics and Fundamental Symmetries'', during which these proceedings were written.

\end{document}